\documentclass[12pt]{article}
\usepackage{color}
\usepackage{latexsym}
\usepackage{amsmath}
\usepackage{amsfonts}
\usepackage{amssymb}
\usepackage{indentfirst} 
\usepackage{tikz}
\usepackage{enumerate}
\usepackage{authblk}

\newcommand{\mR}{{\mathbb R}}

\title{More on superintegrable  models on spaces of constant curvature}
\author[1]{C. Gonera \thanks{e-mail: cezary.gonera@uni.lodz.pl}}
\author[1]{J. Gonera}
\author[2]{J. de Lucas}
\author[1]{W. Szczesek }
\author[2]{B.M. Zawora}

\affil[1]{\small Faculty of Physics and Applied Informatics,
University of Lodz, Pomorska 149/153, 90-236 Łódź, Poland.}
\affil[2]{\small Department of Mathematical Methods in Physics, University of Warsaw, Pasteura 5, 02-093 Warszawa, Poland.}

\date{}
\begin{document}
\maketitle 

\vspace{-0.5cm}

\hfill{\em In memory of Professor Alexey Borisov}

\begin{abstract}
A known general class of superintegrable systems on 2D spaces of constant curvature can be defined by potentials separating in (geodesic) polar coordinates. The radial parts of these potentials correspond either to an isotropic harmonic oscillator or a generalised Kepler potential. The angular components, on the contrary,  are given implicitly by a transcendental, in general, equation. In the present note, devoted to the previously less studied models with the radial potential of the generalised Kepler type, a new two-parameter family of relevant angular potentials is constructed in terms of elementary functions. For an appropriate choice of parameters, the family reduces to an asymmetric spherical Higgs oscillator.

\end{abstract}
\section{Introduction}
\par
All Hamiltonian systems with  $2D$ configuration space and separating in the polar coordinates (geodesic polar ones for curved spaces) admit an additional (apart from the Hamiltonian) independent integral of motion. It corresponds to a separation constant. Under mild assumptions concerning potentials, such systems are integrable. In the generic case of confining potentials, there exists no  further global and functionally independent constant of motion. The system is not superintegrable. Moreover, most of the bounded trajectories, densely covering some rings in the (hyper)plane of motion, are not closed. In the rotationally symmetric case, the Bertrand theorem \cite{1} states that the Kepler and isotropic harmonic oscillator potentials are the only central potentials on the plane for which all bounded trajectories are closed. On the spaces of constant curvature, these potentials are substituted by spherical and pseudo-spherical Higgs oscillator and Schroedinger-Coulomb ones \cite{2,3,4,5,6,7,18}, respectively.

The Bertrand theorem can be generalised for systems with non-central separable potentials defined on spaces of constant curvature: the Euclidean plane, the sphere and the hyperbolic plane \cite{8,9,10}.
The Hamiltonian of these models reads      
\begin{equation}
\label{r0}
H_{k}(r,\varphi , p_{r}, p_{\varphi }) = \frac{p_{r}^{2}}{2} + \frac{1}{s_k^2(r)}\left(\frac{p_{\varphi}^{2}}{2} + c(\varphi)\right)+ a_k (r),
\end{equation}
\;\\\
where $(p_r,p_\varphi )$ stand for the canonical momenta conjugated to (geodesic)polar coordinates $(r,\varphi )$, the functions $s_k^2(r)$ are given by
\begin{equation}
\label{r1}
s_k(r) = 
\begin{cases}
\frac{1}{\sqrt{-k}} \sinh(\sqrt{-k}r), & k < 0,\\
r ,& k = 0,\\
\frac{1}{\sqrt{k}} \sin(\sqrt{k}r), & k > 0,\\
\end{cases}
\end{equation}
\;\\\
where $k$ is the curvature constant while  $a_k (r)$ and $c(\varphi)$  denote  radial and angular potentials, respectively.\\
Under mild assumptions concerning the potentials, the Hamiltonian is explicitly integrable.
Involutive, global and functionally independent Liouville integrals of motion are provided by a generalized ``momentum" $l( \varphi ,p_{\varphi })$ given by

\begin{equation}
\label{r1a}
l( \varphi ,p_{\varphi }) = \frac{p_{\varphi}^{2}}{2} + c(\varphi), 
\end{equation}
\;\\
and the Hamiltonian itself.

Due to Arnold-Liouville theorem, it follows that equations
\begin{equation}
\label{r1b}
\begin{split}
 H_k (r,\varphi , p_{r}, p_{\varphi }) = E,\\ 
 l ( \varphi ,p_{\varphi }) = L, 
 \end{split}
\end{equation}
define, for some intervals of the values of the separation constants $E$ and $L$, a compact and sufficiently regular surface  isomorphic to a 2D torus. On the neighbourhood of this torus, one can introduce action-angle variables. 

In our case, the radial $J^{k}_r$ and angular $J_\varphi $ action variables are given by 

\begin{equation}
\label{r1c}
J^{k}_{r}(E, L) = \frac{1}{\pi }\int_{r_ {min}}^{r _{max}}\sqrt{2(E- a_{k}(r) - \frac{L}{s_k^{2}(r)})}dr,
\end{equation}
 
\begin{equation}
\label{r1cc}
J_{\varphi }(L) = \frac{1}{\pi }\int_{\varphi _{min}}^{\varphi _{max}}\sqrt{2(L - c(\varphi ))}d\varphi,
\end{equation}
 %\begin{equation}
%\label{r1ccc}
%$J_{\varphi }(L) = \frac{1}{\pi }\int_{\varphi _{min}}%^{\varphi _{max}}\sqrt{2(L - c(\varphi ))}d\varphi\\
%\;\\\ 
%J^{k}_{r}(E, L) = \frac{1}{\pi }\int_{r_ {min}}^{r _{max}}%%%\sqrt{2(E- a^{k}(r) - \frac{L}{s_k^{2}(r)})}dr 
%\end{split}
%\end{equation}
%\;\\\\
where $\varphi _{min}$, $\varphi _{max}$, $r_ {min}$, $r _{max}$ are the roots of the relevant integrands. \\
\;\\
It is known that an integrable 2D Hamiltonian allows for three global and functionally independent constants of motion (in other words, it is superintegrable), iff it depends on the linear combination of some action variables with integer coefficients. In our case this means that there exist integers $m,n\in\mathbb{Z}$ such that
\begin{equation}
\label{r1d}
H_k = H_k(mJ^{k}_r + nJ_{\varphi}), 
\end{equation}
or equivalently
\begin{equation}
\label{r1e}
f^k(E) = mJ^{k}_{r}(E,L) + nJ_{\varphi}(L),
\end{equation}
where $f^k(E)$ is a function of energy.

Equation (\ref{r1e}) implies that the integral obtained by taking the partial derivative of   (\ref{r1c}) with respect to the parameter $L$, does not depend on the energy $E$. In fact, this is the condition on the functions $a_k(r)$. One can show \cite{8,9} that it is met if $a_k (r)$ is of oscillator type, namely
\begin{equation}
\label{r2}
 a_k(r) = 
\begin{cases}
 \gamma \mid k\mid \coth^2(\sqrt{-k}r) + \frac{\displaystyle \omega }{\displaystyle \mid k\mid }\frac{\displaystyle 1}{\displaystyle \coth^2(\sqrt{-k}r)}, & \text{if}\;\;k\;<\;0,\\
 \;\\
 \frac{\displaystyle \gamma}{\displaystyle r^2} + \omega  r^2, & \text{if}\;\;k\;=\;0,\\
 \;\\
  \gamma \mid k\mid \cot^2(\sqrt{k}r) + \frac{\displaystyle \omega }{\displaystyle \mid k\mid }\frac{\displaystyle 1}{\displaystyle \cot^2(\sqrt{k}r)}, & \text{if}\;\;k\;>\;0,\\
\end{cases}
\end{equation}
\;\\\
or  $a_k (r)$ is of (generalised) Kepler type
 
\begin{equation}
\label{r3}
 a_k(r) = 
\begin{cases}
 B \mid\! k\!\mid \coth^2{(\sqrt{-k}r)}\! - \!\sqrt{-k}\coth{(\sqrt{-k}r)}\sqrt{D +\! \mid\!k\!\mid F\coth^2{(\sqrt{-k}r)}}, \\
 \;\\
 \frac{\displaystyle B}{\displaystyle r^2} - \frac{\displaystyle \sqrt{Dr^2 + F}}{\displaystyle r^2}, \\
 \;\\
 B\mid\! k\!\mid \cot^2{(\sqrt{k}r)} -  \sqrt{k}\cot{(\sqrt{k}r)}\sqrt{D + \mid \!k\!\mid F\cot^2{(\sqrt{k}r)}}, \\
\end{cases}
\end{equation}
 
In the above formula, the first line corresponds to $k<0$, the second one to $k=0$ and the last one to $k>0$, respectively.

To deal with physically interesting potentials (given by real, single-valued, convex functions with a single minimum) one chooses\\
 $\gamma, \omega, B, D, F \geq 0$ and $ B+ L \pm\sqrt{F} \geq 0 $ as well as $\gamma  + L\geq 0$; in addition, for $k<0$ one puts $\omega> k^2 \gamma$ while $D > 2|k|\sqrt{B^2-F} ( \sqrt{B^2-F} + B)$. 

Knowing the allowed radial potentials $a_k(r)$, one explicitly computes  the corresponding radial actions $J^{k}_{r}(E,L)$.
Inserting them together with the general expression for angular action $J_\varphi(L)$ (given by eqation (\ref{r1cc})) into  (\ref{r1e}) and taking the partial derivative of (\ref{r1e}) with respect to  $L$, we obtain an Abel integral equation for the function $c(\varphi)$. An implicit solution for this equation, representing the
angular potential $c(\varphi )$ on an interval $(\varphi _1, \varphi _2)$ with a single minimum $c_0$ at $\varphi _0 \in (\varphi _1, \varphi _2)$ and two branches, $c_{+} (\varphi )$ for $\varphi  > \varphi _0$ and $c_{-} (\varphi )$ for $\varphi  < \varphi _0$, both going to infinity as $\varphi$ approaches $\varphi _1$ or $\varphi _2$, respectively, is given by a formula
 
\begin{equation}
\label{r4}
 \varphi _{\pm }(c) = \pm \frac{1}{2\nu }(\frac{\pi }{2} - \arcsin{f(c)}) + G(c).
\end{equation}
\;\\
 Here $\varphi _{\pm }(c)$ are the inverse maps of $ c _{\pm }(\varphi)$, $G : R\rightarrow R$ is an arbitrary single-valued function not spoiling the assumed properties (a single valuedness, a single minimum at $\varphi _0$, convex character) of the potential $c(\varphi )$ while the function $f(c)$ reads:
 
\begin{equation}
\label{r5}
 f(c) = 
\begin{cases}
 \frac{\displaystyle 2c_0 +\gamma - c}{\displaystyle c + \gamma },  & \text{for $a_k(r)$ of oscillator type},\\
 \;\ \\
 \frac{\displaystyle c_0 + \sqrt{(c_{0} + B)^2 - F} - c}{\displaystyle \sqrt{(c  + B)^2 - F}}, &  \text{for $a_k(r)$ of (generalized)Kepler type}. \\
\end{cases}
\end{equation}
 
The parameter $\nu $  is given by

\begin{equation}
\label{r6}
 \nu  = 
\begin{cases}
 \frac{\displaystyle 2n}{\displaystyle m},  & \text{for $a_k(r)$ of oscillator type },\\
 \;\ \\
 \frac{\displaystyle n}{\displaystyle m}, &  \text{for $a_k(r)$ of (generalised) Kepler type}, \\
\end{cases}
\end{equation}
\;\\
\;\\ 
A few remarks are in order:
\begin{itemize}
\item If the angular potential $c(\varphi )$ vanishes,then  one deals with models defined by the central potentials $a_k(r)$ given by equations(\ref{r2}) or (\ref{r3}). It appears, that these models describe  superintegrable dynamics provided we put $\gamma =0$ and $B =0=F$. Then, indeed we get the systems allowed by the Bertrand theorem. It means, in particular, that the radial potential of the generalised Kepler type, i.e. with the non-vanishing  parameter $F$, has to be accompanied by a non-trivial angular one.

\item Equations (\ref{r2})--(\ref{r6}) provide the general description of superintegrable systems on 2D configuration spaces of constant curvature with non-central potentials separating in the (geodesic) polar coordinates. However, the angular potentials $c(\varphi )$, unlike the radial $a_k(r)$ ones, are given only implicitly by eq. (\ref{r4}). Explicit solutions for angular potentials $c(\varphi)$ in terms of elementary functions exist only for an appropriate choice of functions $G(c)$. Although there are no general rules or criteria for choosing  $G(c)$ functions, there exist some more or less obvious possibilities. For example, one can simply take $G(c)\!=\!{\rm const}\!=\!\varphi_0$. The resulting angular potential presented already in \cite{8}, for  the radial potentials of the oscillator type reads
\begin{equation}
\label{r7}
c(\varphi) = \frac{\displaystyle 2(\gamma  + c_0)}{\displaystyle \cos2\nu (\varphi  - \varphi _0) + 1} - \gamma,
\end{equation}
 
 and for the radial potentials of the (generalised) Kepler type
 
{\tiny  \begin{equation}
\label{r8}
c_\pm (\varphi)=
\begin{cases}
\frac{\displaystyle J + B\cos^2(2\nu (\varphi  - \varphi _0)) \pm  \cos2\nu (\varphi  - \varphi _0)\sqrt{(J + B)^2 - F\sin^2(2\nu (\varphi  - \varphi _0))}}{\displaystyle \sin^2(2\nu (\varphi  - \varphi _0))} & \varphi  \neq \varphi _0\\
\;\\
c_0 & \varphi  = \varphi _0
\end{cases}
\end{equation}}
 
where $J\equiv c_0 + \sqrt{(c_{0} + B)^2 - F}$. Here $c_{+}(\varphi )$ and $c_-(\varphi)$ correspond to $J+B<0$ and $J+B>0$, respectively.
\end{itemize}

Another, less obvious choices of a $G$ function considered in \cite{10} provides two-parameter families of the angular potentials. In particular, the family corresponding to the radial potentials of the oscillator or Kepler type includes (for a relevant choice of parameters) Poschl-Teller potential. 
Generally, the class of models with harmonic or Kepler radial potentials and their curved counterparts (including, in particular, TTW \cite{11,12} and PW \cite{13} models) has been studied within different approaches in many papers, both on the classical and quantum levels \cite{14}-\cite{29c}.\footnote{Note, that ref.\cite{29c} analyses oscyllator-related and Kepler-related systems on 3D spaces with a constant curvature.}
Most methods of searching for superintegrable systems focus on so called polynomial superintegrability. Assuming from the very beginning the existence of an additional constant of motion polynomial in the momenta, one tries to deduce a form of the potential. This can be done either by solving a relevant differential equation for the potential or by extending in a proper way a Hamiltonian of a lower dimensional integrable system \cite{29a}, \cite{29b}.
However, the systems with the generalised Kepler radial potential have been hardly investigated. It seems that only a few angular potentials compatible with the generalised Kepler one and leading to superintegrable dynamics have been known \cite{8,10}. In addition, they are quite complicated and rather exotic.

The present note is devoted to the study of models with the radial potential of the generalised Kepler type. We introduce a new two-parameter family of $G$ functions which generate the simpler explicit form of the angular potentials. In particular, for an appropriate choice of the parameters, one obtains an asymmetric spherical Higgs oscillator. The superintegrability of the systems is then confirmed by showing that their Hamiltonians (in the action-angle variables) are functions of linear combination (with integer coefficients) of the action variables. In addition, the explicit form of the third (super)constant of motion is presented.

\section{The angular potentials corresponding to the radial potentials of generalised Kepler type}

  We are going to construct explicitly, in terms of elementary functions, the angular potentials $c(\varphi )$ corresponding to the radial  ones $a_k(r)$,
of the generalised Kepler type given by equation (\ref{r3}). To this end, we consider eq. (\ref{r4}) with the function $f$ given by the second line of the formula (\ref{r5}) (where, without losing the generality, we put $c_0=0$) and assume that $G(c)$ satisfies the ansatz\\
\, \\
\begin{equation}
\label{r9}
\cos{2\nu G(c)} = \frac {\alpha c +\beta}{\sqrt{(c+B)^2 - F}},
\end{equation}
where $(\alpha , \beta) \in \mR \times \mR  $ and $(\alpha , \beta) \neq \pm (-1,\sqrt{B^2-F}\equiv J)$. The pairs $(\alpha , \beta) = \pm (-1, J)$ are excluded because they generate functions $G(c)$ breaking the single valuedness of the potentials $c(\varphi)$. The ansatz is consistent provided that for every 
$ c \in [0; \infty )$ one has
\begin{equation}
\label{r10}
\Big \vert \frac {\alpha c +\beta}{\sqrt{(c+B)^2 - F}}\Big \vert \leq 1.
\end{equation}
This finally leads to

\begin{equation}
\label{r11}
|\alpha|\leq 1, \qquad  |\beta|\leq J \qquad \text{and} \qquad
(\alpha , \beta) \neq \pm (-1, J).
\end{equation}

%\begin{equation}
%\label{r11}
%|\alpha|\leq 1 \qquad \text{and} \qquad |\beta|\leq \sqrt{B^2-F}\equiv J.
%\end{equation}

Substituting (\ref{r9}) into (\ref{r4}) one obtains (after some algebraic and trigonometric operations)  the quadratic equation for the potential $c(\varphi)$:
\begin{equation}
\label{r12}
a(\varphi)c^2 +b(\varphi)c + d(\varphi) = 0,
\end{equation}
where
\begin{equation}
\label{r13}
\begin{split}
&a(\varphi) = (\cos{2\nu \varphi} + \alpha)^2,\\
&b(\varphi) = 2(B \cos^2{2\nu \varphi} +(\beta -\alpha J) \cos {2\nu \varphi} +\alpha\beta -(J+B)),\\
&d(\varphi)=(\beta - J\cos{2\nu \varphi})^2.
\end{split}
\end{equation}
 The discriminant of (\ref{r12}):
 \begin{multline}
\label{r14}
\Delta = 4(B+J)\sin^2{2\nu \varphi}\Big( (J-B)\cos^2{2\nu \varphi} +2(\alpha J -\beta)\cos{2\nu \varphi} +(J+B) - 2\alpha\beta \Big) \\\equiv
 4(B+J)\sin^2({2\nu \varphi}) \delta(\varphi)
\end{multline}
is nonnegative for $\alpha , \beta$  satisfying the conditions (\ref{r11}). Then for any $\varphi$ one has two real solutions $c_\pm(\varphi)$ of the form:
{\small \begin{equation}
\label{r15}
c_\pm (\varphi)=\frac{-B\cos^2{2\nu \varphi} + (\alpha J - \beta)\cos{2\nu \varphi} - \alpha\beta +(J+B) \pm \sqrt{B+J}\sin{2\nu \varphi} \sqrt{\delta(\varphi)}}{(\cos {2\nu \varphi} + \alpha)^2}.
\end{equation}}
 
Both solutions are related by the transformation $2\nu \varphi\mapsto 2\pi -2\nu \varphi$, and without loss of the generality we shall use $c_-(\varphi)$, hereafter denoted  by $c( \varphi)$.\\
It can directly be checked that $c(\varphi)$ is the convex function on the interval 
$$(\varphi_1, \varphi_2) = (\frac{1}{2\nu}(\arccos{\alpha} - \pi),\frac{1}{2\nu}(\arccos{\alpha} + \pi))$$  
such that $\lim_{\varphi\rightarrow\varphi_i}c(\varphi)=\infty$ for $i=1,2$. At the point $\tilde{\varphi}\!=\!\frac{1}{2\nu}( - \arccos{\alpha} +\pi)\in (\varphi_1 , \varphi_2)$ both the numerator and denominator of the $c(\varphi)$ go to zero. However, the function $c(\varphi)$ has a well-defined limit there, which can be taken as its value $c(\tilde{\varphi})$. As it should be $c(\varphi)$ attains the unique minimum $c_0=0$ at the point $\varphi_0=\frac{1}{2\nu}\arccos{\frac{\beta}{J}}$. Remarkably, due to the conditions (\ref{r11}), other points, where the first derivative of $c(\varphi)$ vanishes, do not belong to the interval $(\varphi_1 , \varphi_2)$. It follows that the parameter $\alpha$ determines the domain of the function $c(\varphi)$ while $\beta=J\cos{2\nu\varphi_0}$ controls the position of its minimum.
%The plot of the potential $c(\varphi)$ is presented in Fig.1\\
Knowing the meaning of the parameters $\alpha$, $\beta$ one can change them to shape the form of the potential $c(\varphi)$, in particular, to simplify it. For example, putting $\beta=\alpha J$ i.e. requiring the position  $\varphi_0$ of the minimum to be at the midpoint of the interval $(\varphi_1,\varphi_2)$ eliminates terms linear in $\cos{2\nu\varphi}$ and leads to potential 
 
\begin{equation}
\label{15a}
c(\varphi)=\frac{B\sin^2{2\nu\varphi} - \sqrt{F}\sin{2\nu\varphi}\sqrt{\sin^2{2\nu\varphi} + \frac{2(1-\alpha^2)J}{B-J}}}{(\cos{2\nu\varphi}+\alpha)^2}.
\end{equation}
 
Requiring further the minimum of  $c(\varphi)$ to be at $\varphi_0=0$, which is equivalent to taking $\alpha=1$ and $\beta=\alpha J$, yields an asymmetric spherical Higgs oscillator 
 
\begin{equation}
\label{15b}
c(\varphi)= 
\begin{cases}
(B+\sqrt{F})\tan^2{\nu\varphi},  & \varphi \in (-\frac{\pi}{2\nu}, 0),\\
(B-\sqrt{F})\tan^2{\nu\varphi},  & \varphi \in [0,\frac{\pi}{2\nu}).
\end{cases}
\end{equation}
 
Similarly, for $\alpha=-1$, $\beta=-J$, one obtains 
 
\begin{equation}
\label{15c}
c(\varphi)= 
\begin{cases}
(B-\sqrt{F})\cot^2{\nu\varphi},  & \varphi \in (0,\frac{\pi}{2\nu}),\\
(B+\sqrt{F})\cot^2{\nu\varphi},  & \varphi \in [\frac{\pi}{2\nu},\frac{\pi}{\nu} ).
\end{cases}
\end{equation}
This is probably the simplest possible angular potential compatible with the generalized Kepler radial potential that generates the superintegrable dynamics. Note that, the models defined by equations (\ref{r3}), (\ref{15b}) provide a generalization of the Post-Winternitz system \cite{13} (with the symmetric Poschl-Teller potential), which one obtains in the F=0 limit. Putting in addition D=0 in the radial part of the potential one obtains the system considered in the papers \cite{29a}, \cite{29b}.

\section{Superintegrability of the models}
Our models are superintegrable by  construction (see \cite{9} and \cite{10} for details). Nevertheless, this can be independently confirmed by:
\begin{enumerate}[(i)]
\item showing that the Hamiltonian (\ref{r0}), when expressed in terms of action-angle variables, is a function depending on linear combination (with integer coefficients) of the action variables, or
\item
constructing explicitly the additional super constant of motion.

\end{enumerate}

The radial action variables $J^k_r$  read (\cite{10})
\begin{equation}
\label{r17}
J^k_r(E,L) = h^k(E) -\frac{\sqrt{2}}{2}\Big( \sqrt{L+B+\sqrt{F}} + \sqrt{L+B-\sqrt{F}} \Big),
\end{equation}
 where $h^k(E)$ are energy functions related to the relevant radial potential $a_k(r)$ given by (\ref{r3})). The explicit forms of these functions are not important here. 
 
On the other hand, the angular action variables $J_\varphi(L)$,
as sketched in appendix A, are given by the formula 
 
\begin{equation}
\label{r19}
J_\varphi(L) = \frac{\sqrt{2}}{2}\frac{m}{n}\Big( \sqrt{L+B+\sqrt{F}} + \sqrt{L+B-\sqrt{F}} - \sqrt{B+\sqrt{F}} - \sqrt{B-\sqrt{F}} \Big).
\end{equation}
 
Equations (\ref{r17}) and (\ref{r19}) imply that the linear combinations of action variables 
\begin{equation}
\label{r20}
mJ^k_r(E,L) + nJ_\varphi(L), \qquad m,n \in \mathbb{Z},
\end{equation}
are functions of energy only or, equivalently, the Hamiltonian is the function of such combinations, $H_k=H_k(mJ^k_r +nJ_\varphi)$.\\
As mentioned above, this is a necessary and sufficient condition for a system to have the maximal allowed number of constants of motion, i.e. to be maximally superintegrable. In particular, in our case, this implies that the real or imaginary part of the quantity
\begin{equation}
\label{r21}
C^k(r) = g(J^k_r, J_\varphi) \exp{i(m\Psi_\varphi - n\Psi^k_r)},
\end{equation}
where $g(J^k_r, J_\varphi)$ is an arbitrary smooth function of the actions $J_\varphi , J^k_r $, while $\Psi_\varphi , \Psi^k_r $ stand for conjugated angle variables, respectively, provides a global, functionally independent, third constant of motion. The conservation of $C^k$ follows from the fact that its time derivative is proportional to the time derivative of  ``phase" $\phi^k = m\Psi_\varphi - n\Psi^k_r$ ($\dot{C^k} =i\dot{\phi^k}C$), which vanishes due to the Hamiltonian equations of motion for the Hamiltonian 
$$
\mathcal{H}_k(\Psi_\varphi , \Psi^k_r  ,  J_\varphi , J^k_r) =H_k(mJ^k_r +nJ_\varphi).
$$

In the generic case, the final form of phase $\phi^k$ appears to be quite complicated but can be written out explicitly in terms of elementary functions and elliptic integrals. Here we provide the phase $\phi$ for the model defined by the generalised Kepler potential on the plane, i.e. for $k=0$ and asymmetric Higgs angular oscillator (\ref{15b}). It reads
$$
\phi(\varphi,r)=
-\left(2mf_-\arctan\xi_+(\varphi)+n\left(f_-\arcsin\tau_+(r) + f_+\arcsin\tau_-(r)+\frac{\pi}{2}\right)\right)
$$
for $\varphi\in [-\frac{\pi}{2\nu},0]$, while
$$
\phi(\varphi,r)=-\left(2m(f_+\arctan\xi_-(\varphi)-\frac{\pi}{2})+n\left(f_-\arcsin\tau_+(r) + f_+\arcsin\tau_-(r)+\frac{\pi}{2}\right)\right)
$$
for $\varphi\in (0,\frac{\pi}{2\nu})$,
where
\begin{gather*}
 f_\pm \equiv \frac{\sqrt{L\!+\!B\pm\sqrt{F}}}{\sqrt{L+B+\sqrt{F}}+\sqrt{L+B-\sqrt{F}}},\,\, \xi_\pm\equiv \frac{\sqrt{L-(B\pm\sqrt{F})\tan^2{\frac{n}{m}\varphi}}}{(L+B\pm\sqrt{F})\tan{\frac{n}{m}\varphi}}\\\tau_\pm(r) \equiv \frac{\Big( D\mp 2E\sqrt{F}-\frac{2D(L\!+\!B\pm\sqrt{F})}{\sqrt{Dr^2+F}\pm\sqrt{F}}\Big)}{\sqrt{4FE^2+4D(L\!+\!B)E + D^2}}.
\end{gather*} 
%\begin{eqnarray}
% f_\pm \equiv \frac{\sqrt{L+B\pm\sqrt{F}}}{\sqrt{L+B+\sqrt{F}}+\sqrt{L+B-\sqrt{F}}},\,\, \xi_\pm\equiv \frac{\sqrt{L-(B\pm\sqrt{F})\tan^2{\frac{n}{m}\varphi}}}{(L+B\pm\sqrt{F})\tan{\frac{n}{m}\varphi}}\\
%\tau_\pm(r)\equiv \frac{1}{\sqrt{4FE^2+4D(L+B)E + D^2}}\Big( D \mp 2E\sqrt{F}-\frac{2D(L+B\pm\sqrt{F})}{\sqrt{Dr^2+F}\pm\sqrt{F}}\Big).
%\end{eqnarray}
Again, some details of calculations concerning the generic case and relevant definition can be found in Appendix B.

\section{Summary}
This paper deals with 2D superintegrable systems defined on spaces of constant curvature and separating in (geodesic)polar coordinates. There are two classes of such systems. The first one is determined by the radial oscillator potential (or its curved counterparts) and the corresponding angular ones. In the second class, the radial potentials are of generalised Kepler types.\\
In this note, devoted to the analysis of this second class, we have constructed a new two-parameter family of angular potentials compatible with generalised Kepler one and demonstrated that for the appropriate choice of the parameters the family reduces to the spherical asymmetric Higgs oscillator potential.\\
The superintegrability of the systems considered has been verified by showing that their Hamiltonians (in action-angle variables) are functions depending on a linear combination of action variables with integer coefficients as well as by constructing explicitly the additional superconstant of motion.\\
Finally, we would like to remark that our approach can be directly used to constract superintegrable models of spaces of constant curvature and separating in other than geodesic polar coordinates. In principle it can also be applied to study superintegrability of 2D integrable systems with a non-contant curvature presented for instance in \cite{30c}

\section{Acknowledgements}
The authors would like to thank Piotr Kosi\'nski for discussion and reading the manuscript.

J. de Lucas acknowledges funding from the Polish National Science Centre under grant MINIATURA 5 - Nr 2021/05/X/ST1/01797.

The research has been partly supported (C. Gonera and J. Gonera) by University of Lodz grant IDUB 54/2021.

\section*{Appendix A}

A convenient way to compute the action $J_\varphi (L)$ given  in (\ref{r1cc}) is to calculate first its partial derivative
\begin{equation}
\label{r23}
\frac{\partial J_\varphi}{\partial L} =\frac{\sqrt{2}}{2\pi} \int_{\varphi_{min}}^{\varphi_{max}} \frac{d\varphi}{\sqrt{L-c(\varphi)}}.
\end{equation}
Then, an integration and taking into account the condition $J_\varphi(L=c_0=0)=0$ yield the desired explicit formula for $J_\varphi(L)$. The integral on the RHS of equation (\ref{r23}) is non-elementary, in general. To perform it, we substitute
\begin{equation}
\label{r24}
\varphi=\varphi(c)=
\begin{cases}
\varphi_+(c) \quad \text{for}\quad \varphi \geq\varphi_0,\\
\varphi_-(c) \quad \text{for}\quad \varphi < \varphi_0,\\
\end{cases}
\end{equation}
where
\begin{equation}
\label{r25}
\varphi_\pm(c)=\pm\frac{1}{2\nu}\arcsin{f(c)} + G(c),
\end{equation}
and $f(c)$ is given by equation (\ref{r5}), the $G(c)$ functions follows froom equaton (\ref{r9}), while $\nu=\frac{n}{m}$, for $m,n\in\mathbb{Z}$. Then
\begin{equation}
\label{r26}
d\varphi=
\begin{cases}
d\varphi_+(c) \quad \text{for}\quad \varphi \geq\varphi_0,\\
d\varphi_-(c) \quad \text{for}\quad \varphi < \varphi_0,\\
\end{cases}
\end{equation}
where
\begin{equation}
\label{r27}
d\varphi_\pm=\frac{d\varphi_\pm}{dc}dc=\Big( \mp\frac{1}{2\nu}\frac{1}{\sqrt{1-f^2}}\frac{df}{dc} + \frac{dG}{dc}\Big)dc.
\end{equation}
The non-elementary part of the integral (\ref{r23}) proportional to $\frac{dG}{dc}$ vanishes on the interval $(\varphi_{min},\varphi_{max})$ which implies
\begin{multline}
\label{r28}
\sqrt{2}\pi \frac{\partial J_\varphi}{\partial L} = -\frac{1}{\nu}\int_{0}^{L} \frac{1}{\sqrt{L-c}}\frac{1}{\sqrt{1-f^2}}\frac{df}{dc}dc 
\\=\frac{m}{2n}\Big(\frac{1}{\sqrt{L+B-\sqrt{F}}} + \frac{1}{\sqrt{L+B+\sqrt{F}}} \Big),
\end{multline}
for $m,n\in\mathbb{Z}$.
Integrating this equation and using the condition $J_\varphi(\!L\!=\!0)=0$ results in the angle action variable $J_\varphi(L)$ given by (\ref{r19}).

\section*{Appendix B}
The explicit expression for the conserved phase $\phi^k$ can be found using the formula derived in  \cite[see (61)]{10} to give
\begin{equation}
\label{r29}
\phi^k=\frac{\sqrt{2}}{2}m\frac{dL}{dJ_\varphi} \Big(Z(\varphi)-Y^k(r)\Big),
\end{equation}
with $Z(\varphi)$ and $Y^k(r)$ being the following integrals
\begin{equation}
\label{r30}
Z(\varphi)= \int_{\varphi_{min}}^\varphi \frac{d\varphi}{\sqrt{L-c(\varphi)}},
\end{equation}
\begin{equation}
\label{r31}
Y^k(r) = \int_{r_{min}}^r \frac{dr}{s_k^2(r)\sqrt{E-a^k(r)-\frac{L}{s_k^2(r)}}},
\end{equation}
where $a_k(r)$ are potentials given by eqation (\ref{r3}), while
\begin{equation}
\label{r31a}
\frac{dL}{dJ_{\varphi}}= \frac{4n\sqrt{(L+B)^2 - F}}{m\sqrt{2}(\sqrt{L+B+\sqrt{F}} +\sqrt{L+B-\sqrt{F}})}.
\end{equation}

The integrals $Y^k(r)$ are elementary. In particular, for generalised Kepler potential on the plane, one obtains\\
\begin{multline}
\label{r31b}
 Y(r) = \frac{1}{2\sqrt{L + B - \sqrt{F}}}\arcsin{\Big ( \frac{1}{\sqrt{\Delta }}\big (1 + 2\frac{\sqrt{F}}{D}E \big ) - \frac{2(L + B - \sqrt{F})}{\sqrt{Dr^2 + F} - \sqrt{F}}}\Big ) +\\
\;\\
 \frac{1}{2\sqrt{L + B + \sqrt{F}}}\arcsin{\Big ( \frac{1}{\sqrt{\Delta }}\big (1 - 2\frac{\sqrt{F}}{D}E \big ) - \frac{2(L + B + \sqrt{F})}{\sqrt{Dr^2 + F} + \sqrt{F}}\Big )} ,
\end{multline}
with 
$$\Delta  = \displaystyle{1 + 4\frac{E}{D}\left( B + L + \frac{EF}{D}\right)}.
$$

%\begin{equation}
%\label{r32}
%\begin{split}
%Y^0(r)= \frac{1}{2}\Bigg( \frac{1}{\sqrt{L+B+\sqrt{F}}}\arcsin{\frac{1}{\sqrt{4FE^2+4D(L+B)E+D^2}}\big( D-2E\sqrt{F}- \frac{2D(L+B+\sqrt{F})}{\sqrt{Dr^2+F}+\sqrt{F}}\Big) } \\
%&\frac{1}{\sqrt{L+B-\sqrt{F}}}\arcsin{\frac{1}{\sqrt{4FE^2+4D(L+B)E+D^2}} \Big( D+2E\sqrt{F}- \frac{2D(L+B-\sqrt{F})}{\sqrt{Dr^2+F}-\sqrt{F}} \Big) \Bigg) }\\
%& \frac{\pi}{4}\Big( \frac{1}{\sqrt{L+B+\sqrt{F}}} +\frac{1}{\sqrt{L+B-\sqrt{F}}} \Big)\\ 
%\end{split}
%\end{equation}

To  compute  the non-elementary in generic case integral $Z(\varphi)$, i.e. equation \eqref{r30}, for $|\alpha|<1$, $|\beta|<J$, we perform the substitution (\ref{r24}). Then, 
\begin{equation}
\label{r33}
Z(\varphi)= I_1(\varphi) +I_2(\varphi),
\end{equation}
where
\begin{equation}
\label{r34}
\begin{split}
&I_1(\varphi)=\int_{L}^{\tilde{c}=c(\varphi)}\frac{1}{\sqrt{L-c}}\frac{1}{2\nu}\frac{1}{\sqrt{1-f^2}}\frac{df}{dc}dc=\frac{1}{2\sqrt{2}\nu}\frac{1}{\sqrt{B+\sqrt{B^2-F}}}\times\\
&\\
&\times\Big( \frac{B-\sqrt{F} + \sqrt{B^2-F}}{(B-\sqrt{F})(B-\sqrt{F}+L)}\arccos{\frac{\tilde{c}(L+2(B-\sqrt{F})) - L(B-\sqrt{F})}{(\tilde{c}+B+\sqrt{F})L}}+ \\
&\\
&+ \frac{B+\sqrt{F} + \sqrt{B^2-F}}{(B+\sqrt{F})(B+\sqrt{F}+L)}\arccos{\frac{\tilde{c}(L+2(B+\sqrt{F})) - L(B+\sqrt{F})}{(\tilde{c}+B-\sqrt{F})L}}\Big),
\end{split}
\end{equation}
while
\begin{equation}
\label{r35}
\begin{split}
&I_2(\varphi)=\int_L^{\tilde{c}=c(\varphi)}\frac{1}{\sqrt{L-c}}\frac{dG}{dc}dc=\frac{1}{4\nu}\frac{1}{\sqrt{1-\alpha^2}}\times\\
&\\&\times \int_{L+B}^{\tilde{c}+B}\!\!\!\Big(\frac{\beta-\alpha(B-\sqrt{F})}{z-\sqrt{F}} + \frac{\beta-\alpha(B+\sqrt{F})}{z+\sqrt{F}}\Big)\frac{dz}{\sqrt{(L+B-z)(z-z_+)(z-z_-)}}
\\
&\\&=\frac{1}{2\nu}\frac{1}{\sqrt{(1-\alpha^2)L + B -\alpha\beta +\sqrt{(\alpha B -\beta)^2 +(1-\alpha^2)F}}}\times\\
&\\
& \times\Big( \frac{\beta -\alpha(B-\sqrt{F})}{L+B-\sqrt{F}}\Pi (\Lambda,n_+,p) + \frac{\beta - \alpha(B+\sqrt{F})}{L+B+\sqrt{F}}\Pi (\Lambda,n_-,p)\Big),
\end{split}
\end{equation}
where
\begin{equation*}
\begin{gathered}
\Lambda=\arcsin{\sqrt{\frac{(L-\tilde{c})(1-\alpha^2)}{L(1-\alpha^2) +B-\alpha\beta-\sqrt{(\alpha B -\beta)^2 +(1-\alpha^2)F}}}},\\\\
n_\pm=\frac{(1-\alpha^2)L+B-\alpha\beta-\sqrt{(\alpha B -\beta)^2 +(1-\alpha^2)F}}{(1-\alpha^2)(L+B \mp\sqrt{F})}, \\\\
p=\sqrt{\frac{L(1-\alpha^2) +B-\alpha\beta -\sqrt{(\alpha B - \beta)^2 + (1-\alpha^2)F}}{L(1-\alpha^2) +B-\alpha\beta +\sqrt{(\alpha B - \beta)^2 + (1-\alpha^2)F}}},\\\\
z_\pm = \frac{-\alpha(\alpha B-\beta) \pm \sqrt{(\alpha B -\beta)^2 +(1-\alpha^2)F}}{(1-\alpha^2)},
\end{gathered}
\end{equation*}
and $\Pi(\Lambda,n,p)$ represents the elliptic integral of third kind (see \cite{30}) given by
$$
\Pi(\Lambda,n,p)=\int_0^\Lambda\!\!\! \frac{d\alpha}{(1-n\sin^2{\alpha})\sqrt{1-p^2\sin^2\alpha}}= \int_0^{\sin\Lambda}\!\!\!\!\!\!\!\!\frac{dx}{(1-nx^2)\sqrt{(1-px^2)(1-x^2)}}.
$$
%\section{References}

\end{document}